\documentclass[preprint,showpacs,preprintmbers,amsmath,amssymb]{revtex4}

\usepackage{graphicx}% Include figure files
\usepackage{dcolumn}% Align table columns on decimal point
\usepackage{bm}% bold math
\usepackage{subfigure}

\begin{document}

\title{ $\pi \Xi$ Correlations: 
an Interweaving of Resonance Interaction, Channel Coupling and Coulomb Effects.}
\author{B.O. Kerbikov}
\affiliation{Institute for Theoretical and Experimental Physics, 117218, Moscow, Russia}
\author{L.V. Malinina}
\affiliation{M.V. Lomonosov Moscow State University D.V. Skobeltsyn
Institute of Nuclear Physics, 119992, Moscow, Russia }

\date{\today}

\begin{abstract}
A method is presented for analysis of correlation function of two non-identical particles with strong and Coulomb interactions, resonance formation, channel coupling and spin structure. For resonance reactions we derive a formula giving the small distance contribution to the correlation function. The formalism is used to analyze the preliminary RHIC data on  $\pi^{\pm} \Xi^{\mp}$ correlation measurements. 
The $\Xi^{*}(1530)$ resonance is successfully described. 
The $\pi \Xi$ source size is obtained.

\end{abstract}


%\keywords:{ MC hadron generator, freeze-out surface, fluid flow
%velocity}

\maketitle
\section{\label{sec1}Introduction}

Measurements of momentum correlations of two 
low relative momentum particles 
produced in heavy ion collisions 
provide a unique information on spatio-temporal picture of the emission source 
at the level of few fermis.
The study of the collision region on the femtometer scale via 
two particles correlations is called {\it femtoscopy} 
(see, for example, \cite{pod89}-\cite{lis05}).
At the early stages the studies were focused on the 
identical pions correlations arising from the wave function 
symmetrization  \cite{GG}. This type of analysis
has a deep analogy with the ``HBT interferometry'' used in astronomy \cite{HBT}.
Since then, measurements have been performed
for different systems of both identical and non-identical hadrons.
High-statistics data sets were accumulated in heavy ion experiments at 
AGS, SPS and RHIC accelerators \cite{WA97}-\cite{STAR}.
Correlations are significantly affected by the Coulomb and/or strong
final state interaction (FSI) between outgoing particles.
The non-identical particle correlations due to the FSI provide information
not only about space-time characteristics of the emitting source, but also about
the average relative space-time separation between the emission points
of the two particle species in the pair rest frame (PRF) \cite{llen96}.

Maybe the most exotic system studied recently by STAR collaboration 
is $\pi-\Xi$ \cite{Sumbera07,Chaloupka_SQM06}: 
the particles composing the pair have one order of magnitude
difference in mass plus  $\Delta$B=1/$\Delta$S=2 gap in baryon/strangeness 
quantum numbers. 
It is challenging to study 
FSI of such exotic meson-baryon system and to extract information 
about the $\pi-\Xi$ scattering lengths. 
The other important reason to study  $\pi-\Xi$ 
correlations is that
multistrange baryons are expected to 
decouple earlier, than other particle species 
because of their small hadronic cross-sections~\cite{Bass}, 
allowing one to extract the space-time interval between the 
different stages of the fireball evolution.

Preliminary results for the $\pi \Xi$
system are available from STAR Collaboration
\cite{Sumbera07,Chaloupka_SQM06}. The following important observations
were made:
\begin{itemize}
\vspace*{ -.3cm} \item Decomposition of the correlation function
$C(\vec{k})\equiv C(k,\cos\theta,\varphi)$ ($\vec{k}$ is the center-of-mass (cms) momentum),
 from $10\%$
of the most central Au+Au collisions into spherical harmonics, provided
the first preliminary values of $R=(6.7 \pm 1.0)$ fm and
$\Delta_{out}=(-5.6 \pm 1.0)$ fm. The negative value of the shift
parameter $\Delta_{out}$ indicates that the average emission point of
$\Xi$ is positioned more to the outside of the fireball than
the average emission point of the pion.

\vspace*{ -.3cm}\item  In addition to Coulomb interaction present
in previous non-identical particle analysis the $\pi^{+} \Xi^{-}$
correlations at small relative momenta provide sufficiently clear
signal of the strong FSI that reveals itself in a peak
corresponding to the $\Xi^*(1530)$ resonance. The
peak's centrality dependence shows a high sensitivity to the
source size.

\vspace*{ -.3cm}\item 
In \cite{Sumbera07,Chaloupka_SQM06} 
was shown qualitative agreement 
with the model calculations 
in the Coulomb region and overestimation of 
the peak in the $\Xi^*(1530)$-region.

\end{itemize}

It is clear that the problem of  $\pi \Xi$
correlations deserves a thorough theoretical treatment 
which is undertaken in the present
paper. 
We will be concentrated on finding the expression for the $\pi \Xi$
wave function (WF) with Coulomb and strong interactions included.

The wave function enters as a building block into a correlation function (CF): 
\begin{equation}
\label{BK1}
 C(\vec{k}) =\int d\vec{r}S(\vec{r})|\Psi_{\vec{k}}(\vec{r})|^{2} .
\end{equation}
here 
%$\vec{k}$ is a relative momentum of the pair, 
$S(\vec{r})$ is a source function, $\Psi_{\vec{k}}(\vec{r})$ 
is the two final-state particles WF.

The influence of the 
Coulomb interaction can be taken into account following the 
standard procedure described in textbooks.
A model independent approach to strong FSI is absent since the low energy hadron interactions can not be described from the first principles of QCD.
Phenomenological approach to the combined treatment of Coulomb and strong FSI is based on the effective range expansion of the strong amplitude.
In case of $\pi \Xi$ system the problem is rather complicated due to the following factors:
\cite{Pratt-Petriconi}-\cite{LL98}.
% $\cite{Led08}
%\cite{Pratt-Petriconi}$^,$\cite{LL98}$^,$\cite{Led08}$^,LL82,LL98,$\cite{Led08}:
\begin{itemize}
\vspace*{ -.3cm}\item The superposition of strong and Coulomb interactions
\vspace*{ -.3cm}\item The presence of $\Xi^*(1530)$ resonance
\vspace*{ -.3cm}\item The spin structure of the w.f. including spin-flip.
\vspace*{ -.3cm}\item The fact that the $\pi^{+} \Xi^{-}$ state is a superposition of $I=1/2$
and $I=3/2$ isospin states and that $\pi^{+} \Xi^{-}$
state is coupled to the $\pi^{0} \Xi^{0}$  and that the thresholds of the two channels are non-degenerate.
\vspace*{ -.3cm}\item The contribution from inner potential region where the structure of the strong
interaction is unknown.
\end{itemize}

The description of the $\pi \Xi$ correlation is a twofold problem. First of all one has to construct the  $\pi \Xi$ WF with all factors enumerated above included.
Secondly, one has to use a reasonable model for the source function. In the present paper we concentrate on the first task, while the source function is taken in the simplest Gaussian form. 
Limitations of the naive Gaussian model are well known and will be briefly discussed at the end of the paper. 
In \cite{Sumbera07,Chaloupka_SQM06} the Blast Wave model for the source was used in
combination with the FSI model from \cite{Pratt-Petriconi}, the simultaneous description of the Coulomb
and resonance regions was not obtained. Here we obtained successful description of the source size using simple Gaussian source model.
It is clear that more elaborated source model is need to describe simultaneously source sizes and emission asymmetries.

%%%%%%
%my prev To solve the problem of $\pi \Xi$  correlations one
%needs to understand well the FSI in $\pi \Xi$ system
%and influence of the source model on the correlation function.
%We considered here the simplest Gaussian source function
%and obtain a successful description
%of the experimental data. However the realistic source model including
%e.g. source expansion and resonance decays can destroy such description.
%(probably in \cite{Sumbera07,Chaloupka_SQM06} using the Blast-Wave model
%not allowed to obtain simultaneous description of the
%Coulomb and the resonance regions with the FSI model from \cite{Pratt-Petriconi}).

The paper is organized as follows. Section \ref{sec2} is devoted to a somewhat pedagogical introduction
to the description of the FSI.
Our purpose will be to expose the formalism in a way 
suited for the construction of the CF. In Section \ref{sec3}
we consider the spin-isospin structure of the WF and the coupling between charged and neutral
$\pi \Xi$ channels. Here we also present the first fit to the experimental data. This fit encounters 
problem in the resonance region, and the blame for this discrepancy lies in the disregard of the small distance region. 
The small distance contribution to the WF is derived in Section  \ref{sec4}.
Section \ref{sec5} is devoted to the analysis of the experimental data. In  Section \ref{sec6}
the main results are presented and open problems are formulated.

\maketitle
\section{\label{sec2}The basic FSI formalism.}

As it was adumbrated above, the structure of the $\pi \Xi$ WF is rather complicated.
To set the scene for the detailed treatment of the FSI in the   $\pi \Xi$
system we start with a pedagogical introduction into the FSI problem in the 
context of femtoscopy. Almost all results presented in this paragraph may be found in the 
literature. What we have tried here is to gather in one place the properties of the out-state WF
$\Psi_{\vec{k}}^{(-)}(\vec{r})$ needed to calculate the CF. 
Thus our treatment does not seriously overlap with any other in the literature.

There are two complete sets $\Psi_{\vec{k}}^{(\pm)}(\vec{r})$ of the continuous 
spectrum wave functions. 
In $\Psi_{\vec{k}}^{(+)}(\vec{r})$ the ingoing flux is nonzero only in the direction
 $\vec{n}=-\vec{m}$, $\vec{n}=\vec{k}/k$, $\vec{m}=\vec{r}/r$. This corresponds
 to particles moving along $\vec{n}$ toward the center.
 The flux of outgoing particles in $\Psi_{\vec{k}}^{(+)}(\vec{r})$ consists of two
 parts: non-scattered ones moving along $\vec{n}$ in the direction
 $\vec{n}=\vec{m}$ out of the center, and scattered particles moving in all directions. 
On the contrary,
in $\Psi_{\vec{k}}^{(-)}(\vec{r})$ ingoing particles are 
moving from all directions and outgoing particles are propagating only along the ray 
$\vec{n}=\vec{m}$.
Asymptotically at $r \to \infty$
both functions $\Psi_{\vec{k}}^{(\pm)}(\vec{r})$ contain the same plane wave
 $e^{i \vec{k} \vec{r}}$:
\begin{equation}
\label{BK2} 
\Psi_{\vec{k}}^{(+)}(\vec{r}) \sim e^{i \vec{k} \vec{r}}+
f(\vec{n}\vec{m}) \frac{e^{i k r}}{r} 
\end{equation}

\begin{equation}
\label{BK3} 
\Psi_{\vec{k}}^{(-)}(\vec{r}) \sim e^{i \vec{k} \vec{r}}+
f^{*}(-\vec{n}\vec{m}) \frac{e^{-i k r}}{r} 
\end{equation}
Coulomb interaction and channel coupling bring some distortions 
into these equations - see below.
The following relation between $\Psi_{\vec{k}}^{(+)}(\vec{r})$
and $\Psi_{\vec{k}}^{(-)}(\vec{r})$ holds 
\begin{equation}
\label{BK4} 
\Psi_{\vec{k}}^{(-)}(\vec{r})=\Psi_{-\vec{k}}^{(+)*}(\vec{r}) 
\end{equation}
    
The wave function $\Psi_{\vec{k}}^{(-)}(\vec{r})$ which is called the out-state 
is used to describe particles produced in some process. 
Suppose, for example, that some state $\phi(\vec{r})$ is created in the potential $U(\vec{r})$. Then the flow of the particles with the wave vector $\vec{k}$
emitted from the center is given by
\begin{equation}
\label{BK5} 
N(\vec{k})=\int d \vec{r} \phi(\vec{r}) [\Psi_{\vec{k}}^{(-)}(\vec{r})]^{*} .
\end{equation}
The out-state  $\Psi_{\vec{k}}^{(-)}(\vec{r})$ has to be used in Eq.~(\ref{BK1}) 
for the correlation function.

Consider a pair of particles interacting via attractive Coulomb potential (like
 $\pi^{+} - \Xi^{-}$). The WF
$\Psi_{\vec{k}c}^{(-)}(\vec{r})$ reads

\begin{equation}
\label{BK6} 
\Psi_{\vec{k}c}^{(-)}(\vec{r})=e^{\pi \eta/2} 
\Gamma(1+i\eta)e^{i\vec{k}\vec{r}}F(-i\eta,1,-i(kr+\vec{k}\vec{r})). 
\end{equation}

Here $\eta=1/ka_{B}$, $a_{B}$ is the Bohr radius, $a_{B}=1/\alpha \mu$ (=214~fm
for  $\pi^{+} - \Xi^{-}$), $\mu$
is the reduced mass (=126 MeV for $\pi^{+} - \Xi^{-}$),
$\alpha=1/137$,  $F(\alpha, \gamma, z)$ is the confluent hyper-geometric function.
The region $\eta \ge 1$ is called the atomic energy range.
In this region Coulomb interaction dominates even in presence of strong 
interaction. The wave function is normalized so that
\begin{equation}
\label{BK7} 
|\Psi_{\vec{k}c}^{(-)}(\vec{r}=0)|^{2}=\frac{2 \pi \eta}{1-e^{-2\pi\eta}}=c^{2}(k), 
\end{equation}
where $c^{2}(k)$ is Gamow's factor.
  
As an illustration let us take in (\ref{BK1}) a Gaussian model for the source function
\begin{equation}
\label{BK8} 
S(\vec{r})=(8\pi^{3/2}R^{3})^{-1}\exp(-r^{2}/4R^{2}).
\end{equation}
Expanding $\Psi_{\vec{k}c}^{(-)}(\vec{r})$ in $r/a_B$
and keeping only the first order term we obtain the following result
for the correlation function (\ref{BK1}):
\begin{equation}
\label{BK9} 
C(k)=c^{2}(k)\{1-\frac{8}{\sqrt{\pi}} \frac{R}{a_B}+8\frac{R^2}{a_B^{2}}\}.
\end{equation}
The partial wave expansion of the Coulomb WF is given by
\begin{equation}
\label{BK10} 
\Psi_{\vec{k}c}^{(-)}(\vec{r})=\frac{1}{kr}\sum_{l=0}^{\infty}i^{l}(2l+1)e^{-i\sigma_l}
F_l(\eta,kr)P_l\left(\frac{\vec{k}\vec{r}}{kr}\right).
\end{equation}
where $\sigma_l$ is the Coulomb phase
\begin{equation}
\begin{array}{{c}}
\label{BK11}
\sigma_l=arg \Gamma(l+1-i\eta),
\\
e^{-i\sigma_{l}}=\frac{\Gamma(l+1+i\eta)}{|\Gamma(l+1+i\eta)|},
\end{array}
\end{equation}
and $F_l(\eta,kr)$ is the regular solution of the Coulomb problem 
\begin{equation}
\label{BK12} 
F_l(\eta,kr)=N_l e^{-ikr}(kr)^{l+1}F(l+1+i\eta,2l+2,2ikr),
\end{equation}
\begin{equation}
\label{BK13} 
N_l=\frac{2^{l}}{(2l+1)!}\{c^2(k)\prod_{m=1}^{l}( m^2+\eta^2) \}^{1/2}.
\end{equation}
Asymptotically $F(\eta,kr)$ behaves as 
\begin{equation}
\label{BK14}
F_l(\eta,kr)\sim \sin\left(kr-\frac{l\pi}{2}+\eta ln(2kr)+\sigma_l\right).
\end{equation}
Irregular solution is denoted by $G_{l}(\eta,kr)$ and it has the following asymptotes
\begin{equation}
\label{BK15}
G_l(\eta,kr)\sim \cos\left(kr-\frac{l\pi}{2}+\eta ln(2kr)+\sigma_l\right).
\end{equation}

Asymptotically at $\xi =r(1+\cos{\theta}) \gg 1/k$ one has
\begin{equation}
\label{BK16} 
\Psi_{\vec{k}c}^{(-)}(\vec{r})\to\left(1+i\frac{\eta^{2}}{k\xi}\right)
\exp\{i\vec{k}\vec{r}+i \eta ln(k\xi)\}+
\frac{f_c^{(-)}(\theta)}{r}\exp \{-ikr-i\eta ln(2kr)\},
\end{equation}

\begin{equation}
\label{BK17} 
f_c^{(-)}(\theta)=\frac{\eta}{ 2k\cos^{2}{\frac{\theta}{2}} }
\frac{\Gamma(1+i\eta)}{\Gamma(1-i\eta)}
\exp\{-2i\eta ln(\cos{\frac{\theta}{2}})\}.
\end{equation}

When the produced particles interact only via strong forces their wave function
outside the interaction range (at $r > \epsilon \simeq 1$~fm )
has the following form
\begin{equation}
\label{BK18} 
\Psi_{\vec{k}s}^{(-)}(\vec{r})=\frac{1}{kr}\sum_{l=1}^{\infty}i^{l}(2l+1)e^{-i\delta_l}
\chi_{kl}(r)P_l\left(\frac{\vec{k}\vec{r}}{kr}\right).
\end{equation}
The radial wave function $\chi_{kl}(r)$ asymptotically behaves as
\begin{equation}
\label{BK19} 
\chi_{kl}(r)\sim \sin{\left(kr-\frac{l\pi}{2}+\delta_l\right)}
\end{equation}
Consider now a situation when FSI is caused by the
combined action of the Coulomb and strong interactions.
Then at $r> \epsilon$ we have
\begin{equation}
\label{BK20} 
\Psi_{\vec{k}cs}^{(-)}(\vec{r})=\frac{1}{kr}\sum_{l=0}^{\infty}i^{l}(2l+1)
e^{-i(\widetilde{\delta}_l+\sigma_l)}u_{kl}(r)P_l\left(\frac{\vec{k}\vec{r}}{kr}\right).
\end{equation}
The radial WF $u_{kl}(r)$ is a regular solution of 
the Schrodinger equation containing the sum of Coulomb and strong potentials.
An important point is that $\widetilde{\delta}_l$ entering into 
(\ref{BK20}) is not identical with the pure hadronic phase shift $\delta_l$ in (\ref{BK18}).
The difference is called 
the Coulomb correction \cite{BK_1} and will be explicitly introduced later.

At $r>\epsilon$ the radial wave function in  (\ref{BK20})
has the form
\begin{equation}
\label{BK21}
u_{kl}(r)= F_l(\eta,kr)\cos{\widetilde{\delta}_{l}}+G_l(\eta,kr)\sin{\widetilde{\delta}_l}.
\end{equation}

Making use of (\ref{BK21}) and of the expression (\ref{BK10}) for the Coulomb WF we obtain
\begin{eqnarray}
\label{BK22} 
&\Psi_{\vec{k}cs}^{(-)}(\vec{r})  = 
\Psi_{\vec{k}c}^{(-)}(\vec{r})+
\frac{1}{2kr}\sum_{l=0}^{\infty} i^{l}(2l+1)
e^{-i\sigma_l}(e^{-2i\widetilde{\delta}_l}-1) \lbrack F_l(\eta,kr)+iG_l(\eta,kr)\rbrack  
P_l\left(\frac{\vec{k}\vec{r}}{kr}\right)=  &  \nonumber \\ & =
\Psi_{\vec{k}c}^{(-)}(\vec{r})+
\sum_{l=0}^{\infty}\frac{2l+1}{\sqrt{4\pi}}\varphi_{l}(kr)T_{l}(k)P_l\left(\frac{\vec{k}\vec{r}}{kr}\right),&
\end{eqnarray}
where
\begin{equation}
\label{BK23} 
\varphi_{l}(kr)=\frac{\sqrt{4 \pi}}{kr} i^{l} e^{-i \sigma_{l}}
\left(\frac{F_l+iG_l}{i}\right),
\end{equation}
\begin{equation}
\label{BK24} 
T_{l}(k)=\frac{e^{-2i \widetilde{\delta}_{l}}-1}{-2i}
\end{equation}
Recalling the asymptotic form of $\Psi_{\vec{k}c}^{(-)}(\vec{r})$
given by Eqs.~(\ref{BK16})-(\ref{BK17}) and the asymptotic behavior of $F_l$
and $G_l$ defined by Eqs.~(\ref{BK14})-(\ref{BK15}) one arrives at the following
asymptotic of $\Psi_{\vec{k}cs}^{(-)}(\vec{r})$

\begin{equation}
\label{BK25} 
\Psi_{\vec{k}cs}^{(-)}(\vec{r})\sim
\left(1+i\frac{\eta^{2}}{k\xi}\right)e^{i\vec{k}\vec{r}+i\eta ln(k\xi)}+
\left(  f_{c}^{(-)}(\theta)+f_{cs}^{(-)}(\theta)\right)\frac{1}{r}
e^{-i(kr+\eta ln 2kr)},
\end{equation}
 where the Coulomb amplitude $f_{c}^{(-)}(\theta)$
 is given by (\ref{BK17}) and the Coulomb modified strong amplitude 
 $f_{cs}^{(-)}(\theta)$ is equal to 
\begin{equation}
\label{BK26} 
f_{cs}^{(-)}(\theta)=\frac{i}{2k}\sum_{l=0}^{\infty}(-1)^{l}(2l+1)
e^{-2i\sigma_l}(e^{-2i\widetilde{\delta}_l}-1)  
P_l\left(\frac{\vec{k}\vec{r}}{kr}\right).
\end{equation}

Now let us turn to the low energy expansion of the WF (\ref{BK22}).
For the moment we are not interested in its spin-isospin structure. 
This problem will be addressed in the next Section.
Modification of the effective range expansion due to Coulomb 
interaction is a problem whose solution is well 
known \cite{BK_alpha,BK_beta,BK_gamma}.  Here we quote the results 
relevant for the construction of the out-state WF 
$\Psi_{\vec{k}cs}^{(-)}(\vec{r})$.
Coulomb corrections are most important for the $S$-wave. The basic role in the low energy expansion is played by the effective range function $M(k)$.
In $S$-wave we have
\begin{equation}
\label{BK27} 
T_{0}=\frac{e^{-2i \widetilde{\delta}_{0}}-1}{-2i}=\frac{k c^{2}(k)}{M(k)+\Delta+i\widetilde{k}}
\end{equation}
Here $\Delta$ is the so-called Schwinger's correction  \cite{BK_6}
\begin{equation}
\label{BK28}
\Delta=\frac{2}{a_B}\left( ln\frac{2\epsilon}{a_B}+2\gamma \right),
\end{equation}
 $\gamma \simeq 0.577$ is Eugler's
constant, $\widetilde{k}$ is the Coulomb corrected momentum.
\begin{equation}
\label{BK29}
\widetilde{k}=c^2(k)k-\frac{2i}{a_B}h(\eta),
\end{equation}

\begin{equation}
\label{BK30}
\end{equation}
\[ h(x)=ln(x) +Re \Psi\left(1+\frac{i}{x}\right)\simeq \left\{
\begin{array}{ll}
\frac{x^2}{12} & x \ll 1 \\
-\gamma+ln(x)+\frac{1.2}{x^2} & x \gg 1
\end{array} \right. \]

with $\Psi(z)=\frac{d}{dz}ln \Gamma(z)$. 
In what follows the Schwinger's correction will not be taken into account. 
The function $M(k)$
has the usual effective range expansion
\begin{equation}
\label{BK31}
M(k)=\frac{1}{a}+\frac{1}{2}R_0k^{2}+\dots .
\end{equation}
Similar procedure for the $P$-wave leads to the result
\begin{equation}
\label{BK32} 
T_{1}=\frac{e^{-2i \widetilde{\delta}_{1}}-1}{-2i}=\frac{k^{3}(1+\eta^{2})c^{2}(k)}{N(k)+i\widetilde{k}k^{2}(1+\eta^{2})},
\end{equation}
with 
\begin{equation}
\label{BK33}
N(k)=\frac{1}{b}+\frac{1}{2}R_1k^{2}+\dots .
\end{equation}
where $b$ is the scattering volume with the dimension $fm^{3}$, and $R_1=-3/d$
with $d$ having the interpretation 
of the range of forces. The Coulomb correction in 
$P$-wave similar to $\Delta$ 
(see (\ref{BK26})) can be neglected.

The important feature of the $\pi \Xi$ system is the existence of
 the $P$-wave 
resonance $\Xi(1530) P_{13}$ with a width $\Gamma=9.1$~MeV.
 The corresponding value of the c.m. momentum is $k=150$~MeV,
 $\eta=1/k a_B\simeq 1/160$ and therefore Coulomb corrections in this region are small.
Let us recast the $P$-wave term (\ref{BK32}) into the Breit-Wigner form. We get
\begin{equation}
\label{BK34} 
T_{1}=\frac{k^{3}(1+\eta^{2})c^{2}(k)}{N(k)+i\widetilde{k}k^{2}(1+\eta^{2})} \simeq 
-\frac{1}{2}\frac{\Gamma}{E-E_R-i\Gamma/2},
\end{equation}
where 
\begin{equation}
\label{BK35} 
k^2=\frac{\lambda(s,M^2,m^2)}{4s}\simeq 2 \mu E,
\end{equation}
\begin{equation}
\label{BK36} 
E=\sqrt{s}-(M+m),
\end{equation}
\begin{equation}
\label{BK37} 
E_R=M^{*}-(M+m),
\end{equation}
\begin{equation}
\label{BK38} 
\Gamma= \frac{2d}{3\mu}k^{3}
\end{equation}
with $M$ and $m$ being the masses of $\Xi$ and $\pi$ correspondingly,
$M^{*}=1530$~MeV is the mass of the $\Xi^{*}$ resonance,
$\mu= m M (M+m)^{-1}$. The scattering volume $b$ is expressed in terms of the Breit-Wigner resonance as
\begin{equation}
\label{BK39} 
b= \frac{d}{3\mu E_R}.
\end{equation}

\maketitle
\section{\label{sec3}Spin-Isospin Structure of The Wave Function.}

Up to now the spin-isospin structure of the 
$\pi^{\pm} \Xi^{\mp}$ WF has been disregarded. In this section we shall wind up this lacuna.

It is sensible to begin by considering the angular momentum decomposition of the WF with a given isospin $I$. The isospin structure of the WF as well as the coupling to the $\pi^{0} \Xi^{0}$
channel will be included at the next stage. 
 The amplitude of meson-baryon scattering has the following form:

\begin{equation}
\label{BK40}
T=f+g(\vec{\sigma} \vec{n})  
\end{equation}

\begin{equation}
\label{BK41}
\vec{n} = \frac{\lbrack \vec{k},\vec{k^{'}} \rbrack} { |\lbrack \vec{k},\vec{k^{'}} \rbrack|} 
\end{equation}

The second term in (\ref{BK40}) corresponds to the spin-flip amplitude$^1$
\footnotetext[1]
{The importance of the elastic spin-flip and charge-exchange amplitudes 
in this problem was indicated to the authors by R.~Lednicky.}.
Now let us write down the expression for the WF with the spin-flip term included. 
This expression will replace the spin-less equations 
(\ref{BK20}),(\ref{BK22}). For any type of the interaction  
the conserved quantum numbers are $j^{2}$, $l^{2}$
and $j_z=m_j$. The WF having definite values of the above quantum numbers is constructed 
from the direct product of the WFs (\ref{BK20}),(\ref{BK22}) and the spin WFs

\begin{equation}
\nu_1 =  
\left(\begin{array}{c}
1 \\
0
\end{array}   \right),
\nu_2 =  
\left(\begin{array}{c}
0 \\
1
\end{array}   \right)
\label{BK42} 
\end{equation}
corresponding to $m_j=m_s=+1/2$ and  $m_j=m_s=-1/2$ respectively.
 The resulting expressions read

\begin{eqnarray}
& \Psi_{m_j=1/2}^{(-)}  =    \Psi_{\vec{k}c}^{(-)} 
\left(\begin{array}{c}
1 \\
0
\end{array}   \right) 
 + &  \nonumber \\
+ & \sum_{l=0}^{\infty} \phi_{l}(kr) 
\left( 
T_{l+} \sqrt{\frac{l+1}{2l+1}}  
\left(\begin{array}{c}
\sqrt{l+1} Y_{l,0} \\
\sqrt{l} Y_{l,-1}
\end{array}   \right) 
+ T_{l-} \sqrt{\frac{l}{2l+1}}  
\left(\begin{array}{c}
\sqrt{l} Y_{l,0} \\
-\sqrt{2l+1} Y_{l,-1}
\end{array}\right)
\right) &
\label{BK43}
\end{eqnarray}

\begin{eqnarray}
& \Psi_{m_j=-1/2}^{(-)} =  \Psi_{\vec{k}c}^{(-)} 
\left(\begin{array}{c}
0 \\
1
\end{array}   \right)
+ &  \nonumber \\ 
+ & \sum_{l=0}^{\infty} \varphi_{l}(kr) \left( 
T_{l+} \sqrt{\frac{l+1}{2l+1}}  
\left(\begin{array}{c}
\sqrt{l+1} Y_{l,1} \\
\sqrt{l} Y_{l,0}
\end{array}   \right) 
-
T_{l-} \sqrt{\frac{l}{2l+1}}  
\left(\begin{array}{c}
\sqrt{l+1} Y_{l,1} \\
-\sqrt{l} Y_{l,0}
\end{array}   \right)
\right)
.&
\label{BK44}
\end{eqnarray}

Here the symbols ($l\pm$) correspond to $j= l \pm 1/2$. In deriving (\ref{BK43})-(\ref{BK44})
use was made of the relation $Y_{l,m}^{*}(-\vec{k}/k) = (-1)^{l+m}Y_{l,-m}(\vec{k}/k) $.

The low energy region of $ \pi \Xi$
interaction up to the $\Xi^{*}(1530)$
resonance is dominated by $S$- and $P$- waves. Keeping only these two amplitudes
we can rewrite (\ref{BK43}) as follows

\begin{eqnarray}
&\Psi_{m_j=1/2}^{(-)} =  \Psi_{\vec{k}c}^{(-)} 
\left(\begin{array}{c}
1 \\
0
\end{array}   \right) 
+ \varphi_{0}Y_{00}T_{0} 
\left(\begin{array}{c}
1 \\
0
\end{array}    \right)
+
\frac{1}{\sqrt{3}}\varphi_1 Y_{1,0} \left(
2T_{1+}+T_{1-}
\right)
\left(\begin{array}{c}
1 \\
0
\end{array}    \right)
 + &  \nonumber \\
& + \sqrt{\frac{2}{3}}\varphi_1 Y_{1,-1} \left(
T_{1+}-T_{1-}
\right)
\left(\begin{array}{c}
0 \\
1
\end{array}    \right)
.&
\label{BK45}
\end{eqnarray}
Similar expression holds for $m_j=-1/2$. The last term in (\ref{BK45}) corresponds to spin-flip and
this contribution vanishes if $\widetilde{\delta_{l^{+}}}=\widetilde{\delta_{l^{-}}}$. Recalling about isospin doubling we conclude that the WF (\ref{BK45}) 
contains two $S$-wave amplitudes and four $P$-wave ones. To determine six amplitudes from femtoscopic experiments is hardly possible.
To reduce the number of parameters we shall assume that the dominant interaction in $P$-wave occurs in a state with $j=l+1/2=3/2, I=1/2$ containing the $\Xi^*(1530)$  
resonance (the amplitude $T_{1+}$ in (\ref{BK45})). 
Then the number of parameters becomes equal to three. 
The actual number is two since the parameters of the  $\Xi^*(1530)$  resonance 
are known from the experiment. 
There might be one more parameter. The point is that expressions
(\ref{BK20}), (\ref{BK22}), (\ref{BK43}-\ref{BK45}) for the WF correspond to distances $r>\epsilon \sim 1$~fm
where the strong interaction is assumed to vanish. We shall address this problem in the next section and show 
that the contribution of the inner region is important while $\epsilon$
is not a relevant parameter.

We now proceed to discuss the isospin 
structure of the WF and coupling between charged and neutral $\pi \Xi$ channels.
 There are two $\pi \Xi$ 
systems with opposite electric charges: $\pi^+ \Xi^-$ and $\pi^- \overline{\Xi^-}$. 
For definitiveness we shall consider the charged channel $\pi^+\Xi^-$ which is coupled to 
the neutral channel $\pi^0\Xi^0$. The corresponding thresholds are not degenerate with 
the $\pi^+\Xi^-$ threshold being by $\simeq 11$ MeV higher. 
The problem can be treated either in the channel basis $\{|\pi^+\Xi^-\rangle, |\pi^0\Xi^0\rangle\}$, 
or in the isospin basis $\{|I = \frac{1}{2}\rangle, |I=\frac{3}{2}\rangle\}$. 
The relation between the two frames is given by 
\begin{equation}
 \left(\begin{array}{c}
|\pi^+\Xi^-\rangle \\
|\pi^0\Xi^0\rangle   
\end{array}   \right) = 
\left(\begin{array}{cc}
\sqrt{\frac{2}{3}} & \sqrt{\frac{1}{3}} \\
-\sqrt{\frac{1}{3}} & \sqrt{\frac{2}{3}}  
\end{array}   \right)
\left( \begin{array}{c}
|I=\frac{1}{2}\rangle \\ 
|I=\frac{3}{2} \rangle
\end{array}  \right)
\equiv \hat U 
\left( \begin{array}{c}
|I=\frac{1}{2}\rangle \\ 
|I=\frac{3}{2} \rangle
\end{array}  \right).
\label{BK46}
\end{equation}
Strong interaction is diagonal in the isospin basis while we are interested in the channel 
WF \ $|\pi^+\Xi^-\rangle$ (we shall use the subscript $\alpha$ for the quantities corresponding 
to this channel and $\beta$ --- for shose corresponding to $|\pi^0\Xi^0\rangle$). 
Similar problems have been treated previously by several authors 
\cite{BK_beta}, \cite{BK_gamma}, \cite{BK_8}. 
We shall more or less follow the approach proposed by Shaw and Ross \cite{BK_beta}. 
In Section~\ref{sec2} we have introduced the function $M(k) = k\cot \delta$ which 
allows to perform the effective range expansion in presence of the 
Coulomb interaction (see Eqs.(\ref{BK27}-\ref{BK31})). The generalization to the two-channel system reads 
\begin{equation}
\hat T = \hat k^{l+1/2} (\hat M - i\hat k^{2l+1})^{-1}\hat k^{l+1/2},
\label{BK47}
\end{equation}
where
\begin{equation}
\hat k = \left( 
\begin{array}{cc}
k_{\alpha} & 0 \\
0 & k_{\beta} 
\end{array}   \right), \qquad \hat M = \left(
\begin{array}{cc}
M_{\alpha \alpha} & M_{\alpha \beta} \\
M_{\beta \alpha} & M_{\beta \beta} 
\end{array} \right).
\label{BK48}
\end{equation}
Consider first the $S$-wave.  
In single channel case in the scattering length approximation one has $M=1/a$ (see \ref{BK31}).
 In the two-channel case in the isospin basis one has 
\begin{equation}
\hat M^I = \left( 
\begin{array}{cc}
\frac{1}{a_s} & 0 \\
0 & \frac{1}{a_t} 
\end{array}
\right),
\label{BK49}
\end{equation}
where $a_s$ corresponds to $I=1/2$, and $a_t$ --- to $I=3/2$. Transformation to the channel basis reads 
\begin{equation}
\hat M = \hat U \hat M^I \hat U^{-1} = \frac{1}{3}
\left( 
\begin{array}{cc}
\frac{2}{a_{s}} + \frac{1}{a_{t}} & \sqrt{2}\left( -\frac{1}{a_{s}} +\frac{1}{a_{t}}\right) \\
\sqrt{2}\left(-\frac{1}{a_{s}} +\frac{1}{a_{t}}\right) & \frac{1}{a_{s}} + \frac{2}{a_{t}} 
\end{array}
\right),
\label{BK50}
\end{equation}
where the matrix $\hat U$ was introduced in (\ref{BK46}). 
According to (\ref{BK47}) the two-channel generalization of (27) is 
 \begin{equation}
T_{0}  = k_{\alpha} c^2(k_{\alpha})(M_{\beta \beta} + ik_{\beta})d^{-1}, 
\label{BK51}
\end{equation}
\begin{equation}
d = (M_{\alpha \alpha} + i\tilde k_{\alpha}) (M_{\beta \beta} + ik_{\beta}) - M_{\alpha \beta} M_{\beta \alpha}, 
\label{BK52}
\end{equation}
where as before the Schwinger's correction (\ref{BK28}) has been neglected.
 We can now substitute (\ref{BK51}) into (\ref{BK45}) and obtain the 
closed expression for the
$S$-wave component of the WF.
 However, in the two-channel case this would not be a complete answer. 
 One should add the contribution 
from the coupling to the  $|\pi^0 \Xi^0\rangle$ cnannel, i.e.,  a term $\Psi^{(-)} \sim S^*_{\beta \leftarrow \alpha} \Psi_{\beta}^{(-)}$.
The corresponding contribution to (\ref{BK45}) is
 \begin{equation}
\Psi^{(-)}_{m_j=1/2}(\beta \leftarrow \alpha)_{S} = 
\left(\frac{k_{\beta}}{k_{\alpha}}\right)^{1/2} \omega_{0}Y_{00}R_{0}  
\left( \begin{array}{c}
1 \\
0 
\end{array}   \right).
\label{BK53}
\end{equation}
Here 
\begin{equation}
\omega_{l}(k_{\beta}r_{\beta})=
\frac{\sqrt{4 \pi}}{k_{\beta}r_{\beta}} i^{l} e^{-i \sigma_{l}(k_{\alpha})} \hat h_{l}^{(-)}(k_{\beta}r_{\beta}), 
\label{BK54}
\end{equation} 
\begin{equation}
R_{0} = -c(k_{\alpha})(k_{\alpha}k_{\beta})^{1/2} M_{\alpha \beta} d^{-1}, 
\label{BK55}
\end{equation} 
where  $\hat h_{l}^{(\pm)}(z)$ is Hankel function
\begin{equation}
\hat h_{0}^{(\pm)}(z)=e^{\pm iz},~\hat h_{1}^{(\pm)}(z)=(1 \pm i/z)e^{\pm i (z-\pi/2)}  
\label{BK56}
\end{equation} 
We note that $\omega_l(\rho)=\lim_{\eta \rightarrow 0}{\varphi_{l}(\rho)}$, where $\varphi_{l}(\rho)$
was introduced by (\ref{BK23}).

As it was stated above the $S$-wave component of the $|\pi^+ \Xi^- \rangle$ 
WF depends upon two parameters, namely the isospin scattering 
lengths $a_{s}$ and $a_{t}$. 
With our sign convention moderate attraction corresponds to positive signs of the scattering lengths. 
Next we turn to the $P$-wave. We assume 
that the $\Xi^*(1530)$ resonance with $I=1/2$ plays the dominant role
and therefore  we neglect the $P$-wave with $I=3/2$.
The resonance is coupled to both  $\pi^+ \Xi^{-}$ and $\pi^0 \Xi^{0}$
channels. The corresponding amplitude $T_{1+}$  and $R_{1+}$ are evaluated using the matrix 
$\hat U$ (\ref{BK46}).  
Denoting the resonance width in the isospin basis by $\Gamma$ 
we make the transition to the channel basis 
\begin{equation}
\hat U \hat \Gamma \hat U^{-1} = \hat U\left(
\begin{array}{cc}
\Gamma & 0 \\
0 & 0 
\end{array} \right) \hat U^{-1} = \frac{1}{3}\left( 
\begin{array}{cc}
2\Gamma & -\sqrt{2}\Gamma \\
-\sqrt{2}\Gamma & \Gamma 
\end{array} \right).
\label{BK57}
\end{equation}

In line with (\ref{BK34}) we shall omit the Coulomb corrections and write 
\begin{equation}
T_{1+} = -\frac{1}{2}\frac{(\hat U \hat \Gamma \hat U^{-1})_{\alpha \alpha}}{E-E_R - i\frac{\Gamma}{2}}=
-\frac{1}{3}\frac{\Gamma}{E-E_R - i\frac{\Gamma}{2}}, 
\label{BK58}
\end{equation}

\begin{equation}
R_{1+} = -\frac{1}{2}\frac{(\hat U \hat \Gamma \hat U^{-1})_{\alpha \beta}}{E-E_R - i\frac{\Gamma}{2}}=
\frac{1}{3\sqrt{2}}\frac{\Gamma}{E-E_R - i\frac{\Gamma}{2}}.
\label{BK59}
\end{equation}
In writing (\ref{BK59}) we made the approximation
\begin{equation}
(\hat U \hat \Gamma \hat U^{-1})_{\alpha \beta} = -\Gamma \left(\frac{\Gamma_{\alpha}(k_{\alpha})}{\Gamma} \right)^{1/2}
\left(\frac{\Gamma_{\beta}(k_{\beta})}{\Gamma} \right)^{1/2} 
\simeq -\frac{\sqrt{2}}{3}\Gamma(k_{\alpha}).
\label{BK60}
\end{equation}
Two remarks are in order at this point.
First, the amplitudes $T_l$ and $R_l$ are added non-coherently. This will be visualized in the expression for 
$|\Psi^{(-)}|^{2}$ . Second, in order to calculate the CF according with (\ref{BK1}), or a similar equation, 
one has to take $r_{\beta}=r_{\alpha}=r$. This is a reasonable approximation since the reduced masses in 
$\pi^+ \Xi^{-}$ and $\pi^0 \Xi^{0}$
channels are close to each other.

Now we can collect all pieces together and write the expressions for 
$ \Psi_{m_{j}}^{(-)}$ and $| \Psi_{m_{j}}^{(-)} |^{2}$ with Coulomb interaction included
in all partial waves and strong interaction in $S$ and $P$-waves. It is easy to see that
\begin{equation}
| \Psi_{m_{j}=1/2}^{(-)} |^{2} = | \Psi_{m_{j}=-1/2}^{(-)} |^{2}
\label{BK61}
\end{equation}
and therefore 
\begin{equation}
| \Psi^{(-)} |^{2} = \frac{1}{2} \sum_{m_j=\pm 1/2} | \Psi_{m_{j}}^{(-)} |^{2}= | \Psi_{m_{j}=1/2}^{(-)} |^{2}
\label{BK62}
\end{equation}

So, we write
 \begin{equation}
\Psi^{(-)}_{m_j=1/2} = 
A_{\alpha}  
\left( \begin{array}{c}
Y_{00} \\
0 
\end{array}   \right)
+
A_{\beta}  
\left( \begin{array}{c}
Y_{00} \\
0 
\end{array}   \right)
+
\sqrt{\frac{2}{3}}B_{\alpha}  
\left( \begin{array}{c}
\sqrt{2}Y_{10} \\
Y_{1,-1} 
\end{array}   \right)
+
\sqrt{\frac{2}{3}}B_{\beta}  
\left( \begin{array}{c}
\sqrt{2}Y_{10} \\
Y_{1,-1} 
\end{array}   \right),
\label{BK63}
\end{equation}

\begin{eqnarray}
& |\Psi^{(-)}|^{2} = Y_{00} \left( 
|A_{\alpha}|^{2} + |A_{\beta}|^{2} \right)+
\frac{4}{\sqrt{3}}Y_{00}Y_{10} \left( Re A_{\alpha}B_{\alpha}^{*} + Re A_{\beta}B_{\beta}^{*} \right )
+ & \nonumber \\ 
& + \frac{4}{3} Y_{10}^{2} \left( 
|B_{\alpha}|^{2} + |B_{\beta}|^{2} \right)+
\frac{2}{3}|Y_{1 ,-1}|^{2}  \left( 
|B_{\alpha}|^{2} + |B_{\beta}|^{2} \right)
= & \nonumber \\ 
& = \frac{1}{4\pi}\left( 
|A_{\alpha}|^{2} + |A_{\beta}|^{2}+ 
|B_{\alpha}|^{2} + |B_{\beta}|^{2}
\right)
+ \frac{\cos{\theta}}{\pi} \left( Re A_{\alpha}B_{\alpha}^{*} + Re A_{\beta}B_{\beta}^{*} \right )
+ & \nonumber \\ 
& +
 \frac{3\cos^{2}{\theta}}{4\pi} \left( 
|B_{\alpha}|^{2} + |B_{\beta}|^{2} \right).
& 
\label{BK64}
\end{eqnarray}
Here 
\begin{equation}
A_{\alpha}= \sqrt{4\pi} \Psi_{\vec{k}c}^{(-)} + \varphi_{0}T_{0},  ~ B_{\alpha}= \varphi_{1}T_{1+},
\label{BK65}
\end{equation}
\begin{equation}
A_{\beta}=\left( \frac{k_{\beta}}{k_{\alpha}} \right)^{1/2} \omega_0 R_0, ~ 
B_{\beta}=\left( \frac{k_{\beta}}{k_{\alpha}} \right)^{1/2} \omega_1 R_{1+}.
\label{BK66}
\end{equation}
Equation (\ref{BK64}) is a sought for result which should allow to evaluate the CF according to 
(\ref{BK1}). In the next paragraph we shall see that this is still not a complete answer.
 
\maketitle
\section{\label{sec4}Incorporating the inner region.}

With the WF (\ref{BK64}) at hand we can use Eq.(\ref{BK1}) and try 
to fit the experimental data \cite{Sumbera07,Chaloupka_SQM06} on the CF. 
The results are shown in Fig.~\ref{fig:CF1}
by dashed line for the fireball radii $R=7.0$~fm and zero scattering lengths. 
The low momentum region is properly
described. 
However, in the region of the   
 $\Xi^*(1530)$ resonance we see a dip-bump structure instead  of the experimentally observed 
resonant behavior. The wiggly behavior of the calculated CF is explained by the interference
at $\theta \sim \pi$ between $\Psi_{\vec{k}c}^{(-)}$ and the resonant $P$-wave,
see (\ref{BK22}), or  (\ref{BK44})$^{2}$\footnotetext[2]
{The interference at $\theta \sim \pi$ between the incident wave and the scattred one was observed and explained in \cite{Pratt-Petriconi}.}.
Coulomb effects are not important here and for the sake of clarity they can be neglected, the phenomenon is present already 
in the general expression  (\ref{BK3}). Keeping in  (\ref{BK22}) 
the resonant $P$-wave one has 

\begin{equation}
\psi^{(-)}(\theta=\pi) \sim e^{-ikr} \left(1 + \frac{1}{kr} \frac{\Gamma}{E_R-E+i \Gamma/2} \right).
\label{BK67}
\end{equation}
This leads to the interference term
\begin{equation}
|\psi_{\theta=\pi}^{(-)} |^{2} \sim \frac{2}{kr} \frac{\Gamma (E_R-E)}{(E-E_R)^{2}+\Gamma^{2}/4} .
\label{BK68}
\end{equation}
The last expression clearly displays the dip-bump behavior seen in our calculations.
The same result can be derived from the expression (\ref{BK64}) for $|\psi^{(-)} |^{2}$.

Contrary to this result the experimental points shown in Fig.~\ref{fig:CF1}
exhibit a resonant structure. It means that something is missing in our calculations. Recall that the WF-s  (\ref{BK43}) ,  (\ref{BK44}),  
(\ref{BK63}) correspond to the region $r > \epsilon$ where the strong potential is assumed 
to vanish.The contribution from the region $r < \epsilon$ is proportional 
to the time the impinging wave packet spends there. 
For the resonance this time is much larger than is needed to cross the sphere with the radius $r=\epsilon$.

We come now to a formal account of the small distance contribution. First, we split the CF (\ref{BK1})
into two parts
\begin{equation}
\label{BK69}
 C(\vec{k}) \simeq \int_{r>\epsilon} d\vec{r}S(\vec{r})|\Psi_{\vec{k}}^{(-)}(\vec{r})|^{2} +
S(0)\int_{r<\epsilon} d\vec{r}|\Psi_{\vec{k}}^{(-)}(\vec{r})|^{2}.
\end{equation} 

The source function was taken out of the second  integral because $S(\vec{r})$ has a characteristic scale $R \ge 4 fm \gg \epsilon$. 
The integral over the region $r < \epsilon$ entering into the last equation can not be evaluated directly since the structure of 
the WF at small distances is unknown. However this
integral can be expressed in terms of the scattering phase shifts using the Luders-Wigner formula \cite{BK_1,BK_8,BK_9,BK_10}.
This approach is simple and effective when FSI is dominated by a narrow resonance. Otherwise one can apply the method proposed in \cite{LL82} 
(see also \cite{Pratt-Petriconi}).

Consider the state with $j=3/2$, $I=1/2$ containing the  $\Xi^*(1530)$ resonance.
For the radial WF $\chi_{j,I}(r)=\chi_{3/2,1/2}(r)$ the following relation holds
\begin{equation}
\label{BK70}
J = \int_{0}^{\epsilon}dr |\chi_{3/2,1/2}(r)|^{2}=\frac{1}{2k^{2}}\left( \frac{d \delta_{3/2,1/2}}{dk}
+\epsilon -\frac{1}{2k}\sin{2(k\epsilon +\delta_{3/2,1/2}-\pi/2)}\right).
\end{equation} 
The phase is given by
\begin{equation}
\label{BK71}
\delta_{3/2,1/2}=-arctg \frac{\Gamma}{2(E-E_R)}.
\end{equation} 
From (\ref{BK70}) and (\ref{BK71}) we obtain after a short calculation 
\begin{eqnarray}
\label{BK72}
2k^{2}J = \frac{k}{\mu} \frac{\Gamma/2}{(E-R_R)^{2} +\Gamma^{2}/4} + \epsilon
+\frac{\left((E-E_R)^{2}-\Gamma^{2}/4 \right)\sin{2k\epsilon}-\Gamma (E-E_R)\cos{k\epsilon}}{ 2k \left( \ (E-E_R)^{2}+\Gamma^{2}/4 \right)}
\end{eqnarray} 
where $\mu$ is the $\pi \Xi$ reduced mass. 
This expression may be to a high accuracy approximated by the first term only. 
Indeed, at $E=E_R$ the sum of the last two terms yields
\begin{equation}
\label{BK73}
\epsilon -\frac{sin{2k\epsilon}}{2k} \simeq \frac{2}{3}(k\epsilon)^{2}\epsilon \simeq \epsilon,
\end{equation} 
since $(k\epsilon)\sim 1$. The first term at $E=E_R$ is given by
\begin{equation}
\label{BK74}
\frac{2k}{\mu \Gamma} \simeq \frac{2}{\Gamma}\gg \epsilon.
\end{equation} 
Physically (\ref{BK74}) means that particles spend inside the small distance region
the time $\sim 1/\Gamma$ which is much larger than the time $\epsilon/v = \epsilon \mu/k$
needed to cross this region.

Therefore from (\ref{BK70}) and (\ref{BK64}) we obtain for the inner region contribution entering into 
(\ref{BK69}) the following result
\begin{eqnarray}
\label{BK75}
& \int_{r<\epsilon} d\vec{r}|\Psi_{\alpha}^{(-)}(\vec{r})|^{2}=
\frac{2}{3}\int_{r<\epsilon} d\vec{r}|\Psi_{3/2,1/2}^{(-)}(\vec{r})|^{2}= 
\frac{2}{3}\int d \Omega \left( \frac{1+3 \cos^{2}{\theta}}{4 \pi} \right)
\int_0^{\epsilon} dr (4 \pi |\chi_{3/2,1/2}(r)|^{2})= 
 & \nonumber \\ 
& =\frac{16 \pi}{3}\int_0^{\epsilon} dr |\chi_{3/2,1/2}(r)|^{2}=
  \frac{8 \pi}{3k\mu}\frac{\Gamma/2}{(E-E_R)^{2}+\Gamma^{2}/4}.
&
\end{eqnarray} 
Here the factor $2/3$ is the isospin Clebsch-Gordan coefficient squared, the factor $4 \pi$ in front of
$|\chi_{3/2,1/2}(r)|^{2}$ results from our normalization condition, see (\ref{BK23}) and (\ref{BK54}).
We can rewrite this relation in a relativistic form, as 
\begin{eqnarray}
\label{BK76}
 \int_{r<\epsilon} d\vec{r}|\Psi_{\alpha}^{(-)}(\vec{r})|^{2}=
\frac{8 \pi}{3 k \mu_{rel}}\frac{2 s \Gamma}{(s-M_R^{2})^{2}+s\Gamma^{2}},
\end{eqnarray}  
\begin{equation}
\label{BK77}
 \mu_{rel}^{-1}= \frac{2\sqrt{s}(s+M_R^{2})}{s^{2}-(M^{2}-m^{2})^{2}},
\end{equation} 
with $M$ and $m$ being the $\Xi^{-}$ and $\pi^{+}$ masses respectively,
out of the resonance region $(s+M_R^{2})$ in (\ref{BK77}) is replaced by $2s$.

In Fig.~\ref{fig:CF1} by solid line we show the CF calculated according to (\ref{BK69}) 
with the inner region included.

\maketitle
\section{\label{sec5}Analysis of the experimental data.}

We now turn to the comparison of the calculated CF with the experimental data presented 
in  \cite{Sumbera07,Chaloupka_SQM06}. From Fig.~\ref{fig:CF1} we
conclude that the data are fairly well described if we take the fireball radius 
$R=7.7 \pm 0.7$~fm. The fitting was performed by the
minimization of the functional 
$\chi^2 = \sum (CF(k)_{exp}-CF(k)_{model})^{2}/\sigma_{exp}^2$.
The only fitted parameter was $R$, the scattering lengths $a_s$ and $a_t$ were
put equal to zero (see below),
calculations were performed for $\epsilon=0.8$~fm,
but as shown above for a narrow resonance the results are stable with respect to variations of $\epsilon$ at $\epsilon \ll R$.

Interpreting Fig.~\ref{fig:CF1} one should keep in mind that in different momentum intervals the CF yields different physical information. 
Unless $S$-wave scattering lengths are large, which can hardly be expected for
the $\pi \Xi$ system, and unless the fireball radius is anomalously small,
attractive Coulomb interaction becomes dominant already 
at $k \le 0.05$~GeV/c. Strong interaction effects in this region are roughly speaking proportional to $a_{i}^{2}/R^{2}$ $(i=s,t)$ and therefore negligible for $a_{i} \ll R $. 
Therefore a good fit in Fig.~\ref{fig:CF1} with
 $R=7.7 \pm 0.7$~fm is obtained with zero 
 scattering lengths. 
 Only very precise experiments may allow to determine 
 the values of $a_s$ and $a_t$. 
 This conclusion is illustrated by Fig.~\ref{fig:SL} 
 which shows the low momentum CF for different values of $R$ and $a_s$, $a_t$$^{3}$\footnotetext[3]{Recent lattice calculations yield the result $a(\pi^{+}\Xi^{0})=-0.098\pm 0.017$~fm \cite{Torok}}. 
 Given the present accuracy of the data we simply put the scattering
 lengths equal to zero and ignored correlations of their errors with errors of the source size.

Important information on the fireball radius and on the FSI can be inferred from 
the resonance region.
An important point is that the resonance phenomena is very sensitive to the source size. 
This is illustrated by Fig.~\ref{fig:CF2}.
The fit to the resonance effect allows to obtain the source size value rather 
reliably. Based on the data points presented in 
\cite{Sumbera07,Chaloupka_SQM06} we conclude
that the source size is $R=7.7 \pm 0.7$~fm which agrees with the value obtained in 
\cite{Sumbera07,Chaloupka_SQM06} where the  
 low-$k$ , Coulomb dominated part of the $CF$ was selected for fitting,
 excluding the region of the $Ξ^{∗}$ peak.
Unlike \cite{Sumbera07,Chaloupka_SQM06}
we succeeded in describing $\Xi^{*}(1530)$ resonance region.

It should be however noted that the source model which we used
here (Gaussian in PRF) is oversimplified. The first step toward a more realible model of the source function is suggested by the expression (\ref{BK64}) for the WF squared. This expression is tantamount to the expansion in terms of spherical harmonics with $l = 0,1,2$.
Therefore the CF can be decomposed with the $l = 1$ term giving the emission point difference between $\pi$ and $\Xi$. Such analysis was done in \cite{Sumbera07}.
Other important points left beyond our present investigation are the influence of pions produced from resonance decays and non-Gaussian tail of the source function.
Both factors are significant for pion and kaon production in heavy ion collisions  
\cite{Adams05}-\cite{Afanasiev09}. Experimental data on $\pi \Xi$ correlations were analyzed in \cite{Sumbera07,Chaloupka_SQM06} with the account of these effects. 
Roughly speaking they lead to the increase of the source size but the problem has to be carefully investigated. The same is true for the distortion of the CF by the collective flows. As demonstrated in \cite{Verde07} for $d-\alpha$ correlations with resonance formation collective flow may reduce the source size and lower the resonance peak. 
One should, however, keep in mind that direct production of  $\Xi^{*}(1530)$ may compensate this effect.

\maketitle
\section{\label{sec6}Summary.}

We have developed a formalism to evaluate the CF for the system which comprises several features aggravating the analysis. These features are: superposition of strong and Coulomb 
interaction, presence of a resonance, spin structure, channel coupling. 
We have also proposed a new method to incorporate the contribution from 
the small distances where the structure of the interaction is unknown. 
This technique which is legitimate for large enough source size has been successfully 
applied to the preliminary RHIC data on 
 $\pi^{+} \Xi^{-}$ correlation measurements. 
The Gaussian source in the pair reference frame was considered. 
The resulting value of the source size radius is equal to  $R=7.7 \pm 0.7$~fm. 
There is no sensitivity of the data 
to the values of the $\pi^{+} \Xi^{-}$  $S$-wave scattering lengths.
There is a  large sensitivity of the correlation function to the
source size in the region of $\Xi^{*}(1530)$ resonance.
%The sensitivity in the Coulomb region is much lower.
A good description of the $\Xi^{*}(1530)$ resonance region was obtained.  
The developed formalism can be useful for femtoscopy studies of
the systems with the narrow resonances born due to FSI e.g.
$K^{+} K^{-}$ ($K^{+} K^{-} \to \phi$ ).

\begin{acknowledgments}
The authors would like to thank R.Lednicky for numerous stimulating and clarifying discussions and remarks.
We are also grateful to M.Sumbera and P.Chaloupka whose pioneering study of the problem motivated our work.
Remarks from A.Stavinsky and K.Mikhailov are gratefully acknowledged.
The authors are indebted to financial support from grants RFBR-09-02-08037,
RFBR-08-02-92496, NSh-4961.2008.2

%%%%%%%%%%%%%%%%%%%%%%%%%%%%%%%%%%%%%%%%%%%%%%%%%%%%%%%%%%%%%%%%%
\clearpage
% figures
\begin{figure}[h] 
\begin{center} 
\begin{tabular}{c} 
\includegraphics[width=10cm]{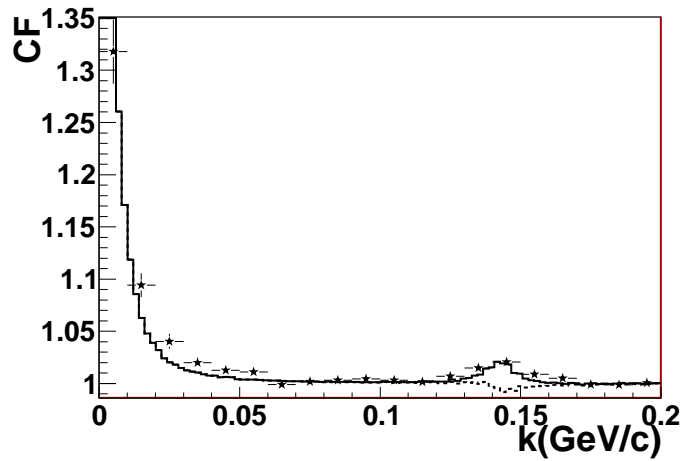} 
\end{tabular}
\end{center} 
\caption{
The CF of $\pi^{+} \Xi^{-}$ system as a function of PRF momentum $k$ 
for $R=7.7$~fm and zero scattering lengths 
without including the inner region correction (dashed line) and
the full calculations (solid line), the STAR collaboration
experimental data points from  
\cite{Sumbera07,Chaloupka_SQM06} (black stars).
\label{fig:CF1}
}
\end{figure}

\begin{figure}[h] 
\begin{center} 
\begin{tabular}{c} 
\includegraphics[width=10cm]{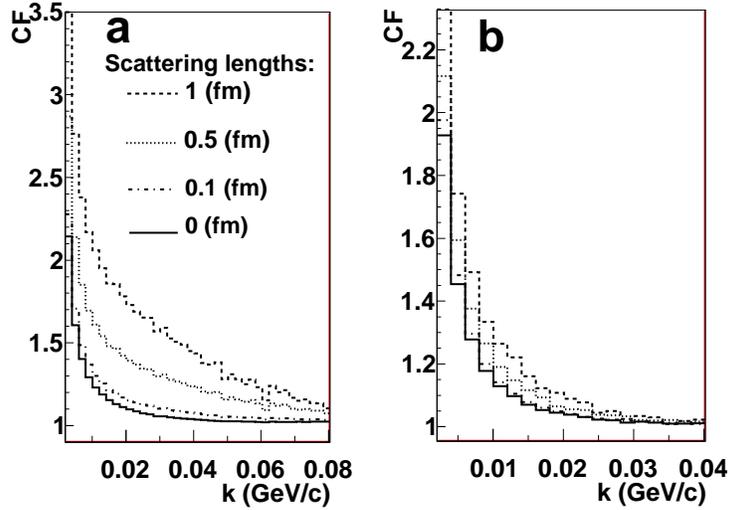}
\end{tabular}
\end{center} 
\caption{
The CF of  $\pi^{+} \Xi^{-}$ system as a function of PRF momentum $k$
for $R=2.0$~fm ({\bf a})
and $R=7.0$~fm ({\bf b}) with the $S$-wave scattering lengths
$a_{\rm 1/2} = a_{\rm 3/2}$: $0$~fm (solid line), 
$0.1$~fm (dashed-doted line), $0.5$~fm (dotted line),
$1.0$~fm (dashed line)
\label{fig:SL}
}
\end{figure}

\begin{figure}[h] 
\begin{center} 
\begin{tabular}{c} 
\includegraphics[width=10cm]{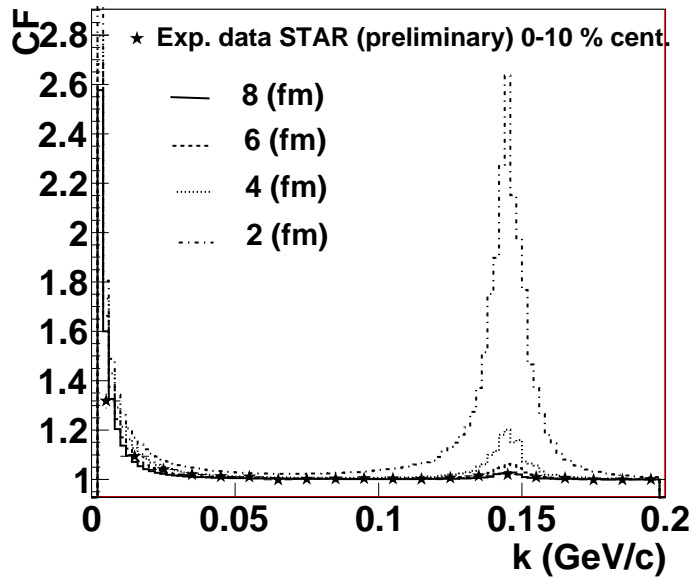} 
\end{tabular}
\end{center} 
\caption{
The CF of  $\pi^{+} \Xi^{-}$ system as a function of PRF momentum $k$
for $R=2.0$~fm (dashed-dotted line),
$4.0$~fm (dotted line), $6.0$~fm (dashed line), $8.0$~fm (solid line)
and the zero scattering lengths (solid line) 
the STAR collaboration experimental data points from 
from \cite{Sumbera07,Chaloupka_SQM06} (black stars).
\label{fig:CF2}
}
\end{figure} 

%%%%%%%%%%%%%%%%%%%%%%%%%%%%%%%%%%%%%%%%%%%%%%%%%%%%%%%%%%%%%%%%%%%

\end{acknowledgments}

%%%%%%%%%%%%%%%%%%%%%%%%%%%%%%%%%%%%%%%%%%%%%%%%%%%%%%%%%%%%%%%%%%%%%%%%%%%%%%%%%%%

\end{document}